\documentclass[11pt,a4paper]{article}

\usepackage{amsmath,amssymb,amsfonts,amsthm}
\usepackage{amsbsy} 
\usepackage{epsfig}
\usepackage{latexsym}
\usepackage[dcu]{harvard}

\begin{document}

\title{Quantum Cosmology}
\author{Steffen Gielen
\\
\\{\tt s.c.gielen@sheffield.ac.uk}
\\
\\{\it School of Mathematics and Statistics, University of Sheffield,}
\\{\it Hicks Building, Hounsfield Road, Sheffield S3 7RH, United Kingdom}
}
\date{\today}

\maketitle

\tableofcontents

\section*{Key points}
\begin{itemize}
\item Canonical quantisation of homogeneous, isotropic cosmology; discussion of ambiguities in this quantisation
\item Construction of explicit solutions, attempts at physical interpretation; conceptual issues (time evolution, unitarity, probability interpretation)
\item Introduction of scalar fields or perfect fluids
\item Discussion of whether classical singularities are resolved
\item Addition of inhomogeneities, possible predictions for primordial cosmology
\item Connections to string theory and loop quantum gravity.
\end{itemize}

\section{Introduction}

Many independent arguments suggest the need for a theory of quantum gravity which unifies the principles of general relativity and quantum mechanics into a coherent framework. Classical general relativity predicts its own demise as a consequence of the Penrose--Hawking singularity theorems (see \citeasnoun{hawkingellis}): in particular, it leads to spacetime singularities inside black holes as well as the Big Bang at the beginning of the entire Universe. A quantum theory of gravity is expected to resolve such singularities, and provide a physical description that avoids the divergences seen classically. This provides a motivation for applying quantum gravity to a homogeneous, isotropic Friedmann--Lema\^{i}tre--Robertson--Walker (FLRW) spacetime with small perturbations, given that such a metric describes the observed Universe to very good precision on large scales. In classical cosmology the assumption of small inhomogeneities allows linearising the equations of general relativity around a highly symmetric background, which makes the resulting equations more tractable than the full Einstein equations. Similar simplifications would be expected in the quantum theory, so that a full understanding of quantum gravity in the nonlinear regime may not be necessary to draw observationally relevant conclusions.

Following this philosophy, the first step in the study of quantum cosmology is to work in the ``minisuperspace approximation'' and develop a quantum theory of FLRW universes. This means passing from a field theory to a finite-dimensional dynamical system, analogous to a system of one or a few interacting particles. In the case most commonly considered, the only degrees of freedom remaining at this level are the scale factor $a$ describing the expansion or contraction of the Universe, and a matter scalar field $\phi$ often taken to represent the inflaton assumed to be the dominant energy component at early times. The universe may have spatial curvature (described by a curvature parameter $k$) and a cosmological constant $\Lambda$ may be present, though this may be absorbed into the scalar field potential energy $V(\phi)$. These variables are subject to a constraint arising from the symmetry under reparametrisations in time. In a Hamiltonian viewpoint, the total Hamiltonian is constrained to vanish. The Hamiltonian may be represented as an operator acting on wavefunctions in the usual fashion, but in trying to make sense of this theory one then faces multiple technical and interpretational issues. To name a few, there may be ambiguities such as ordering ambiguities; there is no unique inner product defining the Hilbert space; given that the Hamiltonian vanishes, there appears to be no time evolution in the system; the dynamical equations have far too many solutions to get insights on the possible initial conditions for the Universe; there is no observer to the Universe, so the standard Copenhagen interpretation of quantum mechanics fails. All of these issues would be faced by a more complete theory of quantum gravity, so one may see quantum cosmology as a simplified case in which they might be addressed.

\section{Minisuperspace models}

The simplest interesting minisuperspace model is obtained by studying pure general relativity with cosmological constant $\Lambda>0$ and closed spatial slices; this could be seen as an approximation to the very early nearly de Sitter Universe during inflation. 

Assuming the line element
\[{\rm d}s^2 = -N^2(t)\;{\rm d}t^2 + a^2(t)\;{\rm d}\Omega^2\,,\]
where ${\rm d}\Omega^2$ is the metric of a homogeneous, isotropic space with constant curvature $k$, the Einstein--Hilbert action (with suitable boundary term) reduces to (see, e.g., \citeasnoun{lorentzianQG})
\begin{equation}
S = 2\pi^2 \int {\rm d}t\;N\left(-3a\frac{\dot{a}^2}{N^2}+3k\,a-\Lambda a^3\right)
\label{einsteinHact}
\end{equation}
where the comoving volume of spatial slices has been chosen to be $2\pi^2$, and units are such that $8\pi G=1$. Varying with respect to the lapse $N$ reproduces the classical Friedmann equation
\[\frac{\dot{a}^2}{a^2N^2}+\frac{k}{a^2}=\frac{\Lambda}{3}\,.\]
The conjugate momentum to $a$ is $p_a=-12\pi^2 \,a\dot{a}/N$, and one obtains the Hamiltonian
\begin{equation}
H=N\left[-\frac{p_a^2}{24\pi^2 a}-2\pi^2(3ka-\Lambda a^3)\right]
\label{GRHamilt}
\end{equation}
subject to the constraint $H=0$.

Following the usual rules of canonical quantisation would amount to replacing $a$ and $p_a$ by operators satisfying $[\hat{a},\hat{p}_a]={\rm i}\hbar$. There are ordering ambiguities in how to present $H$ as an operator, exacerbated by the free choice of lapse $N$, which could classically be taken to be any function of $a$ and $p_a$. One may parametrise these ambiguities in the quantum Hamiltonian to obtain, e.g., (with $N=a$, conformal time)
\begin{equation}
\hat{H}=-\frac{\hat{p}_a^2}{24\pi^2}+{\rm i}\gamma\hat{a}^{-1}\hat{p}_a-U(\hat{a})\,,\quad U(\hat{a})=2\pi^2(3k\hat{a}^2-\Lambda\hat{a}^4)\,.
\label{vilenkinprescr}
\end{equation}
Here $\gamma$ represents the ambiguity from non-commuting operators $\hat{a}$ and $\hat{p}_a$ (see \citeasnoun{vilenkin}), and $U(\hat{a})$ plays the role of a potential for a particle moving in one dimension with coordinate $a$.

An alternative approach is to fix the operator ordering using geometric arguments: for a general model in which all canonical momenta appear in the Hamiltonian quadratically, $H$ is viewed as analogous to the classical Hamiltonian of a particle moving in a curved background, given that it can be written as 
\begin{equation}
H=\frac{1}{2}G^{AB}\pi_A\pi_B+U(q^A)
\label{geometric}
\end{equation}
where $q^A$ may be arbitrary minisuperspace degrees of freedom with conjugate momenta $\pi_A$ (\citeasnoun{Wiltshire}). $G^{AB}$ then defines an inverse (DeWitt) metric on the minisuperspace manifold coordinatised by the variables $q^A$.  An example of a quantisation that is covariant with respect to coordinate transformations on minisuperspace, $q^A \rightarrow Q^B(q^A)$, would be representing $H$ as the differential operator (\citeasnoun{hawkingpage})
\begin{equation}
\hat{H}=-\frac{1}{2}\Box_G + U(q^A) = -\frac{1}{2\sqrt{|G|}}\partial_A\left(\sqrt{|G|}G^{AB}\partial_B\right)+U(q^A)
\label{geometricH}
\end{equation}
acting on wavefunctions in the position representation $\Psi(q^A)$. To extend this notion of covariance to also include redefinitions of the lapse which act as conformal transformations on $G^{AB}$, one can add another term to obtain
\[\hat{H}=-\frac{1}{2}\Box_G - \frac{D-2}{8(D-1)}\mathcal{R} + U(q^A)\]
where $\mathcal{R}$ is the Ricci scalar of the DeWitt metric $G_{AB}$, and $D$ is the dimension of minisuperspace, i.e., the number of degrees of freedom $q^A$ (\citeasnoun{halliwell88}). The operator $-\frac{1}{2}\Box_G - \frac{D-2}{8(D-1)}\mathcal{R}$ on minisuperspace is now conformally invariant. (This proposal is only well-defined for $D>1$.)

After deciding on a choice of operator ordering, the classical constraint $H=0$ is imposed as the quantum Wheeler--DeWitt equation
\[\hat{H}\Psi(q^A)=0\,.\]

For the de Sitter minisuperspace  model which contains only a single minisuperspace variable $a$, the geometric prescription (\ref{geometricH}) depends on the choice of lapse. Taking $N=1/a$ leads to the Wheeler--DeWitt equation
\begin{equation}
\left[\frac{{\rm d}^2}{{\rm d}a^2}-\frac{1}{a}\frac{{\rm d}}{{\rm d}a}-48\pi^4(3ka^2-\Lambda a^4)\right]\Psi(a)=0
\label{specificWDW}
\end{equation}
with general solution
\begin{equation}
\Psi(a)=c_1\,Ai\left(\frac{(12\pi^4)^{1/3}(3k-a^2\Lambda)}{\Lambda^{2/3}}\right) + c_2\,Bi\left(\frac{(12\pi^4)^{1/3}(3k-a^2\Lambda)}{\Lambda^{2/3}}\right) 
\label{WDWgeneralsol}
\end{equation}
where $c_1$ and $c_2$ are constants and $Ai$ and $Bi$ are Airy functions. For other choices of lapse and hence of ordering parameter $\gamma$, analytical solutions can in general only be written in terms of Heun functions.

Having solved this model analytically in terms of possible wavefunctions, one faces the questions of how to choose $c_1$ and $c_2$, and of how to interpret the resulting wavefunction in terms of possible predictions for our Universe. The two questions can be connected when a certain interpretation is invoked to make a choice of solution, as in Vilenkin's tunnelling wavefunction (\citeasnoun{Vilenkin1988})
\[\Psi(a)=Ai(z)+{\rm i}\,Bi(z)\,,\quad z=\frac{(12\pi^4)^{1/3}(3k-a^2\Lambda)}{\Lambda^{2/3}}\,,\]
which satisfies the boundary condition $\lim_{a\rightarrow \infty}\Psi(a)=0$ for $\Lambda<0$: for negative $\Lambda$ there is a classically forbidden region at large $a$, leading to an exponentially growing and an exponentially decaying solution. Vilenkin's boundary condition picks out the decaying one, which is then analytically continued to positive $\Lambda$. At large $a$ and for positive $\Lambda$, one can approximate
\begin{equation}
\Psi_{{\rm V}}(a)\sim \frac{1}{\sqrt{a}} \exp\left(-{\rm i}\,\frac{12\pi^2}{\Lambda}\left(\frac{\Lambda}{3}a^2-k\right)^{3/2}\right)\,.
\label{airy1}
\end{equation}
This is to be contrasted with the other linearly independent solution $\Psi_{{\rm HH}}(a)=Ai(z)$ for which
\begin{equation}
\Psi_{{\rm HH}}(a)\sim \frac{1}{\sqrt{a}} \cos\left(\frac{12\pi^2}{\Lambda}\left(\frac{\Lambda}{3}a^2-k\right)^{3/2}-\frac{\pi}{4}\right)\,.
\label{airy2}
\end{equation}
This second solution is often associated with the no-boundary proposal of \citeasnoun{hartlehawking}, which suggests obtaining a preferred wavefunction from a path integral
\begin{equation}
\Psi(a) = \int \mathcal{D}N\;\mathcal{D}b\;\exp(-I[b,N])
\label{nobound}
\end{equation}
where the integral is to be taken over metrics interpolating between a Euclidean four-sphere (without  boundary) and a Lorentzian spacetime with final boundary condition $b(t_f)=a$, and $I$ is the Euclidean action for gravity evaluated on a spacetime configuration $(b,N)$. Since this action is unbounded from below, the path integral requires deformation of the contour into complex $b$ and $N$ to be well-defined, even if it is initially thought of as Euclidean (\citeasnoun{GibbHawkPerr}).
At leading order in a semiclassical approximation, one then expects (see, e.g., \citeasnoun{whatisnb})
\[\Psi(a)\approx\sum_{{\rm sp}} d_{{\rm sp}}\exp(-I_{{\rm sp}})\]
where ${{\rm sp}}$ labels the saddle points of the path integral, $d_{{\rm sp}}$ are suitable coefficients, and $I_{{\rm sp}}$ is the classical action evaluated on each saddle point. For the de Sitter minisuperspace model one finds four possible saddle points with
\[I_{{\rm sp}}=\frac{12\pi^2}{\Lambda}\left(\pm\,1\,\pm{\rm i}\left(\frac{\Lambda}{3}a^2-k\right)^{3/2}\right)\,,\]
which can (among other possibilities) reproduce the large $a$ asymptotics seen in (\ref{airy1}) and (\ref{airy2}). These saddle points can be expressed as complex solutions to the equations of motion with boundary condition $b(t_i)=0$ and $b(t_f)=a$. For each of them, there is a real contribution to the exponential $\exp(-I)$ which, depending on the choice of signs, can mean small $\Lambda$ values either lead to exponential suppression or enhancement. The last years have seen an active debate on how the coefficients $d_{{\rm sp}}$ might be fixed using a particular choice of integration contour, for instance using Picard--Lefschetz theory as advocated in \citeasnoun{lorentzianQG}. One might also pick out particular saddle points by using an initial boundary condition involving the derivative of the scale factor (\citeasnoun{LoukoHH}).

Having derived a particular solution to the constraint equation (\ref{specificWDW}), one faces the question of how to interpret it. A probability interpretation as in standard quantum mechanics would require a choice of inner product, but the Wheeler--DeWitt equation in general does not suggest a preferred choice.  In the geometric viewpoint leading to (\ref{geometricH}), the Wheeler--DeWitt equation is analogous to a Klein--Gordon equation on a curved manifold, which often has Lorentzian signature. For solutions to such a second-order differential equation, the current
\[J^A = -{\rm i}\sqrt{|G|}G^{AB}\left(\bar\Psi\partial_B\Psi-\Psi\partial_B\bar\Psi\right)\]
satisfies the conservation law $\partial_A J^A = 0$, and hence the integral of $J^0$ (where ``$0$'' denotes a timelike direction in minisuperspace, often the conformal factor of the spatial metric) over a codimension-one hypersurface could serve as a conserved inner product. A common objection against this proposal is that the resulting inner product is not positive definite, and the restriction to solutions with positive norm may be even in principle impossible (\citeasnoun{Kuchar}). In simple models, one can either isolate the solutions with positive norm, or modify the inner product to be positive definite. In the de Sitter minisuperspace model we have considered here, minisuperspace is one-dimensional, and this norm would be proportional to
\[\langle \Psi|\Psi\rangle \propto \frac{{\rm i}}{a}\left(\bar\Psi(a)\,\Psi'(a)-\Psi(a)\,\bar\Psi'(a)\right)\,.\]
One can show that for any solution (\ref{WDWgeneralsol}) this combination is independent of ``time'' $a$, as expected. For any state with positive norm, the norm could be set to one by a constant redefinition, as one would do in usual quantum mechanics.

An alternative proposal is to suggest a Schr\"odinger-like interpretation in which $|\Psi|^2$ directly serves as a probability measure.  This measure is generally not normalisable, and given that one now wants to compare probabilities for all possible configurations of the Universe, including all possible values of variables one might use as a standard of time, does not follow the usual interpretation of quantum mechanics, where one is interested in possible measurement outcomes at a given time. The two issues are closely related, and manifestations of the ``problem of time'' (\citeasnoun{isham}), which plagues all canonical approaches to quantum gravity: the Wheeler--DeWitt equation is a constraint equation rather than describing evolution with respect to a particular time. In the Klein--Gordon probability interpretation, this issue is partially addressed by choosing  a timelike direction on minisuperspace and defining a norm without integrating over this direction. As we saw, such a norm can be finite.

Beyond such structural and mathematical considerations, one may ask what it even means to give probabilities to possible measurement outcomes for a theory that aims to describe the entire Universe. It is clear that a standard Copenhagen interpretation does not apply. One might then investigate alternative interpretations of quantum mechanics, for instance based on decoherent histories (\citeasnoun{DowkerHalliwell}), which may be more suitable in the context of cosmology.

To address the issues with defining a probability measure and the troubles with giving meaning to such a notion, as well as the fact that quantum cosmology minisuperspace models are at best an approximation to full quantum gravity, a common approach is to work only with semiclassical solutions of WKB (Wentzel--Kramers--Brillouin) type. Such solutions can be written as $\Psi=\mathcal{A}e^{{\rm i}S}$ with slowly varying amplitude $\mathcal{A}$ and rapidly oscillating phase $S$. The Klein--Gordon current $J^A$ then approximates to
\[J^A \sim \sqrt{|G|}\,|\mathcal{A}|^2 G^{AB}\partial_B S\]
which flows approximately along classical trajectories, and can be used to define a positive-definite probability measure (\citeasnoun{Wiltshire}). One could argue that notions of probability or unitarity in quantum cosmology are only meaningful in such a semiclassical (WKB) limit. Following this approach, an important question is whether such a semiclassical regime can be reached dynamically; for instance, \citeasnoun{Lehners} showed how this happens for inflationary and ekpyrotic cosmologies, for a choice of state analogous to the Hartle--Hawking no-boundary proposal.

In our initial example, we only studied minisuperspace models for pure gravity, but adding scalar fields is straightforward. With a scalar field $\phi$, the action (\ref{einsteinHact}) becomes
\begin{equation}
S = 2\pi^2 \int {\rm d}t\;N\left(-3a\frac{\dot{a}^2}{N^2}+3k\,a+a^3\frac{\dot\phi^2}{2N^2}-a^3\,V(\phi)\right)
\label{GRscalaract}
\end{equation}
where we have absorbed a cosmological constant into the potential $V(\phi)$. A regime of approximately constant $V(\phi)$ (and $\dot\phi^2/(2N^2)\ll V(\phi)$) can describe slow-roll inflation, so that the previously considered de Sitter minisuperspace model can be seen as the lowest-order approximation to inflation. 

In this more general case, the Hamiltonian becomes
\[H=N\left[-\frac{p_a^2}{24\pi^2 a}+\frac{p_\phi^2}{4\pi^2 a^3}-2\pi^2\left(3k\,a-a^3 V(\phi)\right)\right]\,.\]
Following the Hawking--Page ordering prescription (\ref{geometricH}), the corresponding quantum Hamiltonian would be
\begin{align}
\hat{H}&=\frac{1}{24\pi^2 a^2}\partial_a\left(a\partial_a\right)-\frac{1}{4\pi^2 a^3}\partial_\phi^2-2\pi^2\left(3k\,a-a^3 V(\phi)\right)\nonumber
\\&=\frac{1}{24\pi^2 a}\left(\partial^2_a+\frac{1}{a}\partial_a\right)-\frac{1}{4\pi^2 a^3}\partial_\phi^2-2\pi^2\left(3k\,a-a^3 V(\phi)\right)
\label{WdWwithscalar}
\end{align}
In this case, requiring general covariance with respect to coordinate transformations on minisuperspace fixes the operator ordering uniquely. The superspace metric
\[G_{AB}=\begin{pmatrix} -12\pi^2 a & 0 \cr 0 & 2\pi^2 a^3\end{pmatrix}\]
turns out to be flat; if we take $a \ge 0$, the manifold parametrised by $a$ and $\phi$ with the given metric corresponds to a Milne universe, seen as a part of (1+1) dimensional Minkowski spacetime. This means that the given definition of the Hamiltonian is also covariant under redefinitions of the lapse.

The Wheeler--DeWitt equation $\hat{H}\Psi=0$, for a general choice of potential $V(\phi)$, will not admit any analytical solutions, but there are particular cases for which solutions are known. For instance, one could pass to Cartesian-like coordinates on minisuperspace
\[T=a^2\cosh(\sqrt{2/3}\phi)\,,\quad X=a^2\sinh(\sqrt{2/3}\phi)\,,\]
and redefine the lapse as $N=\bar{N}/a=\bar{N}/(T^2-X^2)^{1/4}$, bringing  (\ref{GRscalaract}) into the form
\[S = 2\pi^2 \int {\rm d}t\;\frac{3\bar{N}}{4}\,\left(-\frac{\dot{T}^2}{\bar{N}^2}+\frac{\dot{X}^2}{\bar{N}^2}+4k-\frac{4\sqrt{T^2-X^2}}{3}V\left(\frac{T}{X}\right)\right)\,.\]
So far this is just a redefinition of variables which is a symmetry of the formalism as we have described it, but these coordinates now suggest particularly simple forms for the potential $V(\phi)$, such as
\[V\left(\frac{T}{X}\right)=\frac{\alpha T + \beta X}{\sqrt{T^2-X^2}}=\alpha\cosh(\sqrt{2/3}\phi)+\beta\sinh(\sqrt{2/3}\phi)\]
which makes the potential linear in terms of $(T,X)$. The resulting Wheeler--DeWitt equation admits explicit solution in terms of Airy functions (\citeasnoun{Garayetal}). For more general potentials, only approximate solutions in terms of a WKB-like expansion $\Psi=\mathcal{A}e^{{\rm i}S/\hbar}$ for small $\hbar$ can be found.

While the majority of the literature is focused on coupling scalar fields due to their interest for inflation or alternatives such as the ekpyrotic scenario, other types of matter can be considered as well. For instance, the action of a perfect fluid in a closed FLRW universe may be defined following \citeasnoun{Brown} as
\begin{equation}
S=\int {\rm d}t\;\left(u\,\dot\chi-2\pi^2 a^3 N\,\rho\left(\frac{u}{2\pi^2 a^3}\right)\right)\,,
\label{brownfluid}
\end{equation}
using a total particle number $u$ and its conjugate variable $\chi$. The energy density $\rho$ is a given function of the particle number density $u/(2\pi^2 a^3)$, encoding the equation of state of the fluid. Different formulations exist, for example following the Schutz formalism (\citeasnoun{Alvarenga:2001nm}), but these often lead to equivalent results given that a perfect fluid in an FLRW universe is only characterised by an energy density $\rho$ given as a fixed function of the scale factor $a$: the Hamiltonian receives an extra term
\begin{equation}
H_{{\rm fl}}=N\frac{M}{a^{3w}}
\label{fluidHamilt}
\end{equation}
where $M$ is a constant of motion (in the case of dust with $w=0$, this would be the total mass of the perfect fluid particles), and $w$ is the equation of state parameter defining the pressure $p=w\rho$. A special case of this construction is the cosmological constant arising as an integration constant in unimodular gravity, which could be seen as a perfect fluid with $w=-1$. The quantity $M$ enters linearly, so if it is seen as a conserved momentum the Hamiltonian is no longer of the general form (\ref{geometric}) in which momenta always appear quadratically. 

The linear momentum $M$ defines a total energy: the Hamiltonian constraint $H=0$ implies that $M$ is equal to a given function of the other dynamical variables. Likewise, the conjugate variable to $M$ (obtained from the variables $u$ and $\chi$ in (\ref{brownfluid})) can be used to define a notion of time evolution, and thus circumvent the problem of time by transforming the Wheeler--DeWitt equation into an equation of Schr\"odinger type.

In the case of dust, the total Hamiltonian obtained from adding (\ref{fluidHamilt}) to (\ref{GRHamilt}) is 
\[H=N\left[-\frac{p_a^2}{24\pi^2 a}-2\pi^2(3ka-\Lambda a^3)+M\right]=N\left[-H_p+M\right]\]
in terms of the physical Hamiltonian
\[H_p=\frac{p_a^2}{24\pi^2 a}+2\pi^2(3ka-\Lambda a^3)\]
which can now be quantised as in the usual Schr\"odinger equation. For instance, in the flat case $k=0$ the resulting three cases for $\Lambda$ correspond to a free particle, harmonic oscillator and inverted harmonic oscillator (\citeasnoun{Husain}).

A subtle point in the quantisation of all minisuperspace models concerns the choice of boundary conditions: minisuperspace has nontrivial boundaries, in particular where the scale factor $a$ vanishes,  as already discussed in \citeasnoun{DeWitt}. The evolution problem defined by the Wheeler--DeWitt equation in general requires specifying boundary conditions at $a=0$, at the point of the classical singularity. We already mentioned a few possible choices for the simplest de Sitter model, in the context of the Vilenkin tunnelling wavefunction and the Hartle--Hawking no-boundary wavefunction. If a dynamical scalar field is present, we saw that minisuperspace has the geometry of the Milne universe, which again has a boundary; the kinetic operator $-\frac{1}{2}\Box_G$ is not naturally self-adjoint. In models where one uses a perfect fluid as a Schr\"odinger time, a requirement of unitary time evolution leads to boundary conditions in order for the quantum (physical) Hamiltonian $\hat{H}_p$  to be represented as a self-adjoint operator.  Unitarity with respect to different possible time variables in general leads to inequivalent types of boundary conditions \cite{dependsonclock}.

It should be clear from this discussion that a single classical cosmological FLRW model can lead to a variety of inequivalent quantum models and preferred solutions in terms of a wavefunction $\Psi$, depending on what requirements are put on the theory or its solutions.

\section{Singularity resolution}

As mentioned in the introduction, classical cosmology suffers from an initial Big Bang singularity which one would like to see resolved by quantum effects. Quantum cosmology models, which explicitly deal with the quantum dynamics of a scale factor $a$, should offer a perspective on whether such resolution can be achieved, and under which conditions.

In the literature one can find a number of different definitions of what it would mean for the cosmological singularity to be resolved. The pioneering paper of \citeasnoun{DeWitt} suggests that the wavefunction should vanish on the singular boundary of superspace (i.e., of the space of 3-geometries), suggesting an interpretation of this as vanishing probability for encountering a singularity. In minisuperspace models, this DeWitt condition becomes $\Psi(a)=0$, and is often considered as a criterion for resolution of the classical singularity. Implicitly, such a criterion assumes an interpretation in which $|\Psi|$ is related to a notion of probability, and as we have discussed above, such interpretation may not be justified in general. The criterion is also not covariant with respect to the conformal transformations related to lapse redefinitions, and analogy with the quantum hydrogen atom suggests that requiring a vanishing wavefunction at the classical singularity may not be necessary (\citeasnoun{Kiefer2019}).

Alternative criteria for singularity resolution use observables that one would consider singular in the classical theory, such as the classically divergent energy density of matter or the classically vanishing spatial volume of the Universe. A strong requirement, particularly put forward in loop quantum cosmology (see the review \citeasnoun{AshtekarSingh}, and our brief discussion below), is that such observables should be represented by bounded operators in the quantum theory. In this case, no measurement in any state could ever result in, e.g., a divergent energy density. It is a main result of loop quantum cosmology that the energy density of matter is indeed bounded from above near the Planck scale. Such an interpretation requires an inner product in which the spectrum of observables can be calculated, and might be considered as too strong given that it makes no reference to the quantum state.

A weaker requirement would be to demand that for any given state, all expectation values of classically singular physical observables remain bounded: for instance, for a classical model which encounters a vanishing volume $a^3\rightarrow 0$ at some finite time, one could ask that the evolution of the quantum state is well-defined everywhere with $\langle a^3 \rangle \ge C^V_\Psi > 0$ for some (in general state-dependent) constant $C^V_\Psi$, or similarly $\langle\rho\rangle \le C^\rho_\Psi < \infty$ for the matter energy density. The important point here is that $C^V_\Psi$ is bounded away from zero, given that minisuperspace only includes the half-line $a\ge 0$ and hence any state will always have a positive $\langle a^3\rangle$. This criterion does not necessarily state that the probability of a singularity is zero, which is what the previous two criteria suggest. It also appears to depend on the choice of time variable used to define evolution (\citeasnoun{dependsonclock}).

One limitation of minisuperspace models is that they are formulated in terms of the classical metric variables, and do not allow for the possibility of the spacetime metric being fundamentally replaced by different degrees of freedom, as suggested by quantum gravity approaches such as string theory. One might speculate that only an embedding of quantum cosmology into a more complete quantum gravity framework would give insights on how classical singularities are resolved. The issue is hence often not very explicitly addressed in more agnostic approaches in which even notions of probability are only assumed to be meaningful at the semiclassical WKB level.

\section{Beyond minisuperspace}

The truncation to exact FLRW spacetimes is a very drastic restriction of the full dynamics of general relativity, both at the classical and quantum level. In cosmology, it only gives the lowest-order  approximation to our Universe, and one would like to add inhomogeneities at least perturbatively.

The Hamiltonian formalism we have introduced in the FLRW case is just a simple version of the Arnowitt--Deser--Misner (ADM) canonical formalism for full general relativity. Indeed, the Einstein--Hilbert action with Gibbons--Hawking--York boundary term coupled to a scalar field $\varphi$,
\[S=\int {\rm d}^4 x\sqrt{-g}\left(\frac{R}{2}-\Lambda-\frac{1}{2}g^{\mu\nu}\partial_\mu\varphi\partial_\nu\varphi-V(\varphi)\right)+\int{\rm d}^3 x\sqrt{h}\,K\]
(where the first integral is over spacetime, $R$ is the Ricci scalar of the spacetime metric $g_{\mu\nu}$ and the second integral is a boundary integral involving the spatial metric $h$ and extrinsic curvature $K$), can be written in Hamiltonian form as
\begin{align}
S &= \int {\rm d}^4 x\left(\pi^{ij}\dot{h}_{ij}+\pi_\varphi\dot{\varphi}-N\,\mathcal{H}-N^i\,\mathcal{H}_i\right)\,,\nonumber
\\ \mathcal{H} & = 2\mathcal{G}_{ijkl}\pi^{ij}\pi^{kl}-\sqrt{h}\left(\frac{{}^3R}{2}-\Lambda\right)+\frac{\pi_\varphi^2}{2\sqrt{h}}+\frac{\sqrt{h}}{2}h^{ij}\partial_i\varphi\partial_j\varphi+\sqrt{h}V(\varphi)\nonumber\,,
\\ \mathcal{H}_i & = -2h_{ik}\nabla_j\pi^{kj}+\pi_\varphi\,\partial_i\varphi\nonumber
\end{align}
(see, e.g., \citeasnoun{Wiltshire}). Here
\[\mathcal{G}_{ijkl}=\frac{1}{2\sqrt{h}}\left(h_{ik}h_{jl}+h_{il}h_{jk}-h_{ij}h_{kl}\right)\]
is known as the DeWitt metric, which can be used to define a metric geometry on superspace (here the space of all 3-geometries, without FLRW restriction). Superspace equipped with this metric has a rich geometric and topological structure, as discussed in \citeasnoun{Giulini}.

Variation with respect to the lapse $N$ and shift $N^i$ leads to the Hamiltonian constraint $\mathcal{H}=0$ and diffeomorphism constraints $\mathcal{H}_i=0$. Attempting to follow a canonical quantisation, we would require that operator versions of these classical constraints annihilate a physical ``wavefunction of the Universe'',
\[\hat{\mathcal{H}}\Psi[h_{ij},\varphi]=0\,,\quad \hat{\mathcal{H}}_i\Psi[h_{ij},\varphi]=0\,.\]
One may write down formal expressions for these operators, using a replacement rule
\[\pi^{ij}(x)\rightarrow -{\rm i}\frac{\delta}{\delta h_{ij}(x)}\,,\quad \pi_\varphi(x)\rightarrow -{\rm i}\frac{\delta}{\delta\varphi(x)}\]
in terms of functional derivatives with respect to the field arguments; notice that $\Psi$ would now need to be a functional on the space of all 3-metrics $h_{ij}$ and scalar field configurations $\varphi$. The resulting equations remain formal as they would require multiple functional derivatives at the same point in space, and there is no rigorous definition for a Hilbert space associated to the functional $\Psi[h_{ij},\varphi]$, from which one could extract notions of probability. Moreover, one would expect the space of solutions to be far too large for a predictive theory, given that already the solution space of classical general relativity is too large to be under control.

An intermediate regime between the minisuperspace truncation and the unknown of full canonical quantum gravity is obtained by linearising around an FLRW spacetime, similar to what is done in classical cosmology. Such an approach might lead to insights for cosmological perturbations, in particular regarding their initial conditions (e.g., for inflation). A general framework for quantum cosmology perturbed around a closed FLRW universe was developed in \citeasnoun{HalliwellHawking}.

In the most immediate application to early universe cosmology, one is only interested in scalar perturbations. There is always a gauge issue in that one may consider perturbations of both the metric and the scalar field around some FLRW background quantities, but only some combination of these quantities is gauge-invariant and hence physical (\citeasnoun{BrandenbergerReview}). A popular gauge is one in which the perturbations only appear in the scalar field,
\[\varphi(x,t)=\phi(t)+\delta\varphi(x,t)=\phi(t)+\sum_n f_n(t)Q_n(x)\,,\]
with the spatial metric unperturbed, i.e., exactly homogeneous and isotropic. $Q_n$ is a suitable complete basis set for scalar perturbations, such as spherical harmonics for the closed FLRW universe, labelled by a general index $n$.

For the minisuperspace background, we obtained the Hamiltonian in (\ref{WdWwithscalar}). Setting $a=e^\alpha$, this can be written as
\[\hat{H}=\frac{1}{24\pi^2}e^{-3\alpha}\left[\partial_\alpha^2-6\partial_\phi^2-48\pi^4\left(3ke^{4\alpha}-e^{6\alpha}V(\phi)\right)\right]\,.\]
The form given in \citeasnoun{HalliwellHawking} is obtained by redefining the dynamical variables by suitable constants, choosing $k>0$ and restricting to chaotic inflation with $V(\phi)=\frac{1}{2}m^2\phi^2$:
\[\hat{H}_0=\frac{1}{2}e^{-3\alpha}\left[\partial_\alpha^2-\partial_\phi^2-e^{4\alpha}+e^{6\alpha}m^2\phi^2\right]\,.\]
In this approach the total Wheeler--DeWitt equation is then given by
\[\left(\hat{H}_0+\sum_n\hat{H}_n\right)\Psi = 0\,,\quad \hat{H}_n=\frac{1}{2}e^{-3\alpha}\left[-\partial_{f_n}^2-e^{4\alpha}+\left((n^2-1)e^{4\alpha}+m^2 e^{6\alpha}\right)f_n^2\right]\,.\]
One now thinks of $\Psi=\Psi(\alpha,\phi;\{f_n\})$ as a wavefunction depending both on the background quantities and all the perturbation variables $f_n$. In this approximation, the perturbation modes $f_n$ are all decoupled, which justifies a product ansatz
\[\Psi(\alpha,\phi;\{f_n\})=\Psi_0(\alpha,\phi)\prod_n\Psi_n(\alpha,\phi;f_n)\]
so that each $\Psi_n$ depends only on a single perturbation variable $f_n$. With this ansatz, the Wheeler--DeWitt equation becomes (\citeasnoun{Kiefer87})
\[\frac{\hat{H}_0\Psi_0}{\Psi_0}-\frac{1}{2}e^{-3\alpha}\left(\sum_{n}\frac{\nabla_A\nabla^A\Psi_n}{\Psi_n}+\sum_{n\neq m}\frac{\nabla_A\Psi_m\nabla^A\Psi_n}{\Psi_m\Psi_n}+2\frac{\nabla_A\Psi_0}{\Psi_0}\sum_n\frac{\nabla^A\Psi_n}{\Psi_n}\right)+\sum_n\frac{\hat{H}_n\Psi_n}{\Psi_n}=0\]
where $\nabla_A:=(\partial_\alpha,\partial_\phi)$ and the index is raised with $\eta^{AB}={\rm diag}(-1,1)$. The first term only depends on the background variables $(\alpha,\phi)$, so must be equal to some function $f(\alpha,\phi)$. One then assumes that the contribution of the sum over $n\neq m$ is negligible compared to the other terms, namely that the terms for different $n$ do not add coherently. Hence one is left with
\[-\frac{1}{2}e^{-3\alpha}\left(\nabla_A\nabla^A\Psi_n+2\frac{\nabla_A\Psi_0}{\Psi_0}\nabla^A\Psi_n\right)+\hat{H}_n\Psi_n=E_n(\alpha,\phi)\Psi_n\]
for each $n$, with some undetermined $E_n(\alpha,\phi)$ such that $\sum_n E_n = f$. A common assumption is that $\Psi_0$ solves the background Wheeler--DeWitt equation (i.e., corresponds to a wavefunction for an exact FLRW solution), in which case the $E_n$ can be taken to be zero, but one can also include backreaction from the perturbations to the background.

The next step is then to assume that the background is described by a semiclassical wavefunction of WKB form $\Psi_0=\mathcal{A}e^{{\rm i}S}$ with rapidly oscillating phase. Then
\[\frac{\nabla_A\Psi_0}{\Psi_0}\approx {\rm i}\nabla_A S\]
and defining $e^{-3\alpha}\nabla_A S\,\nabla^A=:\frac{\partial}{\partial t}$ in terms of a semiclassical WKB time $t$ finally transforms the Wheeler--DeWitt equations into
\[-\frac{1}{2}e^{-3\alpha}\nabla_A\nabla^A\Psi_n-{\rm i}\frac{\partial}{\partial t}\Psi_n+\hat{H}_n\Psi_n=E_n(\alpha,\phi)\Psi_n\,.\]
After the further approximation that derivatives of $\Psi_n$ with respect to the background variables $(\alpha,\phi)$ are small compared to $\nabla_A S$, the first term is dropped and one ends up with an effective Schr\"odinger equation for the perturbation wavefunctions $\Psi_n$,
\[{\rm i}\frac{\partial}{\partial t}\Psi_n=\left(\hat{H}_n -E_n\right)\Psi_n\,,\]
with Hamiltonian given by the perturbation mode Hamiltonian $\hat{H}_n$ possibly corrected by a backreaction term $E_n$. This constitutes the lowest-order semiclassical approximation to cosmological perturbations in quantum cosmology, and the resulting Schr\"odinger equation (which is equivalent to the usual quantum harmonic oscillator) may now be solved to find wavefunctions for each mode $Q_n$. The approximation is somewhat similar to the Born--Oppenheimer approximation in standard quantum mechanics. Following this method and going to the next order in the approximation, one finds a corrected Schr\"odinger equation, including what one may argue are the leading quantum-gravitational corrections to the usual treatment of cosmological perturbation theory, as in \citeasnoun{Kiefer11}.

As for minisuperspace models, one then faces the question of how to choose the quantum state, given that even at this level of approximation the resulting Schr\"odinger equation for each $n$ has infinitely many solutions. There is now a well-defined ground state for each $n$, which will lead to the standard Bunch--Davies vacuum of quantum field theory in curved spacetime, but one might want to see a justification from quantum cosmology for why perturbation modes should start out in this ground state. It turns out that prescriptions for a preferred wavefunction for the FLRW background, such as the Vilenkin tunnelling wavefunction or the Hartle--Hawking no-boundary proposal, can be extended to the inhomogeneous degrees of freedom to define a vacuum state for those. The general claim in the literature is that such a boundary condition will pick out the Bunch--Davies vacuum state for inhomogeneities (see, e.g., \citeasnoun{Laflamme} and \citeasnoun{Vachaspati}), and hence connect the framework of inhomogeneous quantum cosmology to quantum field theory in curved spacetime. It should be said that these general claims have  more recently been disputed by \citeasnoun{nosmooth} who find an inverse Gaussian distribution, i.e., instabilities for quantum inhomogeneities, when using the Vilenkin tunnelling wavefunction for the FLRW background.

Apart from inhomogeneites, one might also want to include anisotropies into a cosmological model. For a homogeneous but anisotropic (Bianchi) model, the methods we have shown for FLRW models can be applied straightforwardly, leading again to a reduced minisuperspace action whose dynamical variables are only functions of time, and to a Wheeler--DeWitt equation after canonical quantisation. One can ask the same questions we have asked before, such as whether classical singularities are resolved, whether there is a consistent probability interpretation, or how one should view time evolution. In the simplest case of a Bianchi I model, it is well-known that the dynamics of anisotropy degrees of freedom written in suitable (Misner) variables is equivalent to those of two free massless scalar fields in a flat FLRW background: writing the metric as
\[{\rm d}s^2 = -N^2\,{\rm d}t^2+V^{2/3}\left(e^{2(\beta_+ +\sqrt{3}\beta_-)}\,{\rm d}x_1^2+e^{2(\beta_+ -\sqrt{3}\beta_-)}\,{\rm d}x_2^2 +e^{-4\beta_+}\,{\rm d}x_3^2\right)\,,\]
the Hamiltonian becomes
\[H = N\left[\frac{3V}{8}\left(-p_V^2+\frac{p_+^2 + p_-^2}{9V^2}\right)\right]\]
and the $\beta_+$ and $\beta_-$ variables are now equivalent to free scalar fields. Hence, this case largely mirrors what we have already discussed. The notation used here is taken from the discussion of notions of time in a vacuum Bianchi I model in \citeasnoun{Malkiewicz}.

\section{Connection to quantum gravity}

Quantum cosmology models are often agnostic to what the fundamental theory of quantum gravity will be, given that they only start with a straightforward canonical quantisation of the Einstein--Hilbert action, ignoring issues of renormalisation, possible higher-curvature corrections, or new degrees of freedom that would become relevant around the Planck scale. One could see this as a strength of the formalism given the absence of a fully established theory of quantum gravity in which one could calculate interesting cosmological dynamics, but it means that quantum cosmology requires many choices and approximations whose validity is not always clear. It hence natural to ask whether lessons learned from (various approaches to) quantum gravity could be used to inform the construction of improved quantum cosmology models.

An important example for this line of research is loop quantum cosmology, where one applies the quantisation methods of loop quantum gravity (LQG) to cosmological models. LQG itself is based on a canonical formalism for general relativity similar to the ADM formalism reviewed above: instead of the 3-metric $h_{ij}$, one works with a triad $e_i^I$ such that $h_{ij}=\delta_{IJ}e^I_i e^J_j$ and defines a new (Ashtekar--Barbero) connection variable
\[A_i^I = \Gamma_i^I + \beta K_i^I\,,\]
where $\Gamma_i^I$ is the metric-compatible, torsion-free (Levi-Civita) connection associated to $e_i^I$, $K^I_i$ is the extrinsic curvature one-form and $\beta$ is a free parameter (known as the Barbero--Immirzi parameter). One can show that $A_i^I$ is canonically conjugate to the inverse densitised triad or ``electric field''
\[E^i_I=\frac{1}{6}\epsilon^{ijk}\epsilon_{IJK}e^J_j e^K_k\,.\]
Hence, general relativity can be defined in terms of an action
\[S=\int {\rm d}^4 x\left(\dot{A}^I_i E_I^i - \Lambda^I G_I - N^i H_i - N H\right)\]
where variation with respect to the lapse $N$ and shift $N^i$ gives the Hamiltonian and diffeomorphism constraints in analogy to the metric case, and $G_I$ are a set of three new constraints (usually called Gauss constraint) corresponding to the new local ${\rm SU}(2)$ gauge invariance of the theory. This viewpoint on general relativity leads to a theory with structural similarities to Yang--Mills theory, which allow the construction of a kinematical Hilbert space on which $H_i$ and $G_I$ may be solved.

The most commonly studied loop quantum cosmological model is the flat FLRW universe (\citeasnoun{AshtekarSingh}). For this model, one chooses the Ashtekar--Barbero connection and electric field as
\[A_i^I = \tilde{c}\,{\bf e}_i^I\,,\quad E^i_I=\tilde{p}\,(\det {\bf e})\,{\bf e}_I^i\]
where ${\bf e}_i^I$ is a given fiducial triad such that $\delta_{IJ}{\bf e}_i^I{\bf e}_j^J$ defines a flat spatial metric, and ${\bf e}_I^i$ is the associated inverse co-triad. These variables correspond to the variables of the metric theory, written in cosmic time with $N=1$, via $\tilde{c}=\beta\dot{a}$ and $\tilde{p}=a^2$. The fundamental Poisson bracket of the theory is
\[\{\tilde{c},\tilde{p}\}=\frac{\beta}{3V_0}\]
where $V_0=\int {\rm d}^3 x\,|\det{\bf e}|$ corresponds to the coordinate volume of spatial slices (or a truncation of space to a ``fiducial cell''). To remove the dependence of the Poisson bracket on $V_0$, one then introduces rescaled variables $c=V_0^{1/3}\tilde{c}$ and $p=V_0^{2/3}\tilde{p}$. For a model describing gravity coupled to a massless scalar field $\phi$, one finds the Hamiltonian
\[H=N\left[\frac{p_\phi^2}{2|p|^{3/2}}-\frac{3}{\beta^2}\sqrt{|p|}\,c^2\right]\,.\]
So far, this is very similar to the minisuperspace models in metric variables that we described previously (with the particular choices $k=V(\phi)=0$). However, one now uses one of the key ingredients of the LQG quantisation: in LQG, the connection $A_i^I$ is not represented as an operator on the Hilbert space. Instead, only finite holonomies and parallel transports along edges of graphs embedded in the space manifold exist. At the level of the cosmological model, assuming the same property for a modified quantisation means that only $e^{{\rm i}\mu c}$ with $\mu\neq 0$ can be defined as an operator, and the limit $\mu\rightarrow 0$ is not well-defined. The expression $c^2$ in the Hamiltonian constraint (which corresponds to the curvature or connection at a point) must be replaced by a holonomy around a small closed loop, leading to a replacement
\[c^2 \rightarrow \frac{1}{\mu^2}\sin^2(\mu\,c)\]
where $\mu$ is determined by the coordinate length of the sides of the loop, which can be a constant or itself a function of dynamical variables. A physically well-motivated choice (\citeasnoun{AshtekarSingh}) is $\mu\propto |p|^{-1/2}$; with this choice, the coordinate (co-moving) length of the sides of the fundamental loop scales as $1/a$ with the expansion of the universe, meaning that the physical discreteness scale is kept constant at the Planck scale. This choice then motivates working with the variables
\[b=\frac{c}{\sqrt{|p|}}\,,\quad v=4|p|^{3/2}{\rm sgn}(p)\,,\quad \{b,v\}=2\beta\]
such that $b$ corresponds to the Hubble parameter $\dot{a}/a$, and $v$ is proportional to a volume $a^3$ times an orientation factor ${\rm sgn}(p)$. The combination $\mu\,c$ is now proportional to $b$. With a particular choice of lapse and fixing other ambiguities, one obtains the loop Wheeler--DeWitt equation
\[\partial_\phi^2\Psi(b,\phi)=\frac{3}{2}\left(\frac{\sin(\lambda b)}{\lambda}\partial_b\right)^2\Psi(b,\phi)\]
where $\lambda$ is now just a numerical constant corresponding to the discreteness scale set by the LQG quantisation. This should be compared with
\begin{equation}
\partial_\phi^2\Psi(b,\phi)=\frac{3}{2}\left(b\partial_b\right)^2\Psi(b,\phi)
\label{WdWscalar}
\end{equation}
in usual quantum cosmology (written in connection variables), which is evidently the $\lambda\rightarrow 0$ limit of the LQG-corrected theory.

As in usual quantum cosmology, the scalar field $\phi$ can now be used as a relational clock, leading to a notion of unitary time evolution for states in a physical Hilbert space that is constructed in analogy with the Hilbert space of full LQG. The introduction of a finite $\lambda$ then changes the properties of the resulting quantum cosmology model quite substantially, and one can show that the classical Big Bang singularity is resolved: the expectation value $\langle \hat{V}(\phi)\rangle$ reaches a finite minimum for any state, rather than asymptotically going to zero as it would in classical cosmology or in a quantum cosmology model based on (\ref{WdWscalar}). Moreover, the energy density of the scalar field $\phi$ has a universal upper bound (valid in any state) near the Planck scale. The resulting model can be extended to include perturbative inhomogeneities, again leading to LQG corrections compared to the usual treatment based on quantum field theory in curved spacetime.

There are many proposals for how to connect quantum cosmology to string theory, either incorporating insights from string theory into quantum cosmology models or using quantum cosmology calculations (seen as approximations to full quantum gravity) to address open questions in string theory. For instance, various ways in which the no-boundary proposal (\ref{nobound}) can be connected to string theory are discussed in \citeasnoun{LehnersReview}. 

One idea is to relate the no-boundary proposal to the AdS/CFT correspondence of string theory, using complex redefinitions of the saddle points to the gravitational path integral. Instead of describing these saddle points as a Euclidean four-sphere transitioning to Lorentzian de Sitter space, one finds an equivalent representation in terms of Euclidean hyperbolic space (with negative definite metric) transitioning via a region with complex metric into Lorentzian de Sitter. (The Einstein equations $R_{\mu\nu}=\Lambda g_{\mu\nu}$ are not invariant with respect to the transformation $g_{\mu\nu}\rightarrow -g_{\mu\nu}$; the left-hand side is invariant whereas the right-hand side gets an extra minus sign. This provides the required mapping between the sphere and hyperbolic space.) The initial Euclidean region is now similar to a saddle-point solution discussed in the context of (Euclidean) AdS/CFT. One feature of such saddle-point solutions is that their action can diverge, corresponding to the infinite volume of hyperbolic space. In AdS/CFT, such divergences can be regulated by adding suitable counterterms to the action, but the interpretation of these counterterms in terms of a boundary quantum field theory has not been fully understood. In this Lorentzian de Sitter interpretation, the same counterterms come from the action evaluated along the complex part of the no-boundary saddle-point solution. For more details on this proposal see \citeasnoun{HertogHartle}.

It is also possible to construct minisuperspace models for gravitational theories different from pure general relativity, and in particular for theories including the additional degrees of freedom (and extra dimensions) of string theory. As an example, consider the work of \citeasnoun{Gasperini}. Starting from  the $(d+1)$-dimensional effective action from string theory
\[S=-\frac{1}{2\lambda_s^{d-1}}\int {\rm d}^{d+1} x\,\sqrt{|g|}e^{-\phi}\left(R+g^{\mu\nu}\partial_\mu\phi\partial_\nu\phi-\frac{1}{12}H_{\mu\nu\alpha}H^{\mu\nu\alpha}+V(\phi)\right)\]
where $\phi$ is the dilaton, $H_{\mu\nu\alpha}$ is the field strength associated to an antisymmetric tensor $B_{\mu\nu}$, and $\lambda_s$ is the fundamental string length, one assumes spatial homogeneity leading to a reduced action
\[S=-\frac{\lambda_s}{2}\int{\rm d}t\,e^{-\bar\phi}\left(\dot{\bar{\phi}}^2+\frac{1}{8}{\rm Tr}\left(\dot{M}{(M^{-1})}^\cdot\right)+V\right),\]
with
\[\bar\phi=\phi-\log\sqrt{|g|}\,,\quad M=\begin{pmatrix} g^{kl} & -{B^k}_j \cr {B_i}^l & g_{ij}-{B_i}^m B_{mj}\end{pmatrix}\,.\]
The matrix representation of the fields $g_{ij}$ and $B_{ij}$ (with indices raised and lowered by $g_{ij}$ and its inverse) is helpful in making manifest a global $O(d,d)$ symmetry of the model, acting as $M\rightarrow\Omega^T M \Omega$.

One can now apply the standard methods of canonical quantisation and derive a Wheeler--DeWitt equation using geometric arguments to fix ordering ambiguities.  Again, one finds that minisuperspace can be interpreted as a flat manifold, which limits the possible ambiguities in defining the quantum Hamiltonian. In the simplest isotropic case where $g_{ij}=a^2\delta_{ij}$ and the $B$ field is absent, and introducing a new time coordinate by ${\rm d}t=e^{-\bar\phi}\,{\rm d}\tau$, the Hamiltonian becomes
\[\hat{H} = \frac{1}{2\lambda_s}\partial_{\bar\phi}^2 - \frac{2d}{\lambda_s}\left(a\partial_a+a^2\partial_a^2\right)+\frac{\lambda_s}{2}V e^{-2\bar\phi}\]
which differs from our previously studied FLRW model for gravity with a scalar field mainly in the form of the potential for the effective scalar degree of freedom $\bar\phi$. Solutions to the corresponding Wheeler--DeWitt equation $\hat{H}\Psi=0$ can be used to study a quantum transition between different classical solutions in the string theory inspired pre-big-bang scenario.

\section{Conclusions}

In the continued absence of a complete theory of quantum gravity, quantum cosmology models provide a starting point for addressing some mathematical and conceptual questions of how quantum mechanics and relativity combine to give us new insights into the origin of our Universe. The standard methods of canonical or path integral quantisation can be straightforwardly applied to minisuperspace models with finitely many degrees of freedom, leading in the canonical setting to the Wheeler--DeWitt equation as the fundamental implementation of dynamics. One then faces questions of interpretation, and has to make choices regarding the notion of Hilbert space, definition of probabilities, or criteria for resolving classical singularities. These issues are dealt with somewhat more straightforwardly if one restricts to a semiclassical WKB-like regime, assuming a particular form of the wavefunction and staying away from high-curvature regimes near the classical singularity. This semiclassical approach has been applied successfully in particular to cosmological inhomogeneities. Partial insights we have from quantum gravity, such as the holonomy quantisation of loop quantum gravity, additional degrees of freedom and extra dimensions in string theory, or holography and the AdS/CFT correspondence, can be incorporated into quantum cosmology models and can change their properties and predictions substantially. In this sense, quantum cosmology can be seen more as a general framework for incorporating quantum mechanics into the dynamics of spacetime in cosmology, rather than a single clearly defined theory. Any ``predictions'' for properties of the Universe that we might extract from quantum cosmology are then also inherently model-dependent.

\bibliography{harvard}

\end{document}